# Time-frequency analysis of microwave signals based on stimulated Brillouin scattering

Dong Ma[a,b], Pengcheng Zuo[a,b] and Yang Chen[a,b,*]

[a] Shanghai Key Laboratory of Multidimensional Information Processing, East China Normal University, Shanghai 200241, China
[b] Engineering Center of SHMEC for Space Information and GNSS, East China Normal University, Shanghai 200241, China
[*] ychen@ce.ecnu.edu.cn

**ABSTRACT**
A novel photonic approach to the time-frequency analysis of microwave signals is proposed based on the stimulated Brillouin scattering (SBS)-assisted frequency-to-time mapping (FTTM). Two types of time-frequency analysis links, namely parallel SBS link and time-division SBS link are proposed. The parallel SBS link can be utilized to perform real-time time-frequency analysis of microwave signal, which provides a promising solution for real-time time-frequency analysis, especially when it is combined with the photonic integration technique. A simulation is made to verify its feasibility by analyzing signals in multiple formats. The time-division SBS link has a simpler and reconfigurable structure, which can realize an ultra-high-resolution time-frequency analysis for periodic signals using the time segmentation and accumulation technique. An experiment is performed for the time-division SBS link. The multi-dimensional reconfigurability of the system is experimentally studied. An analysis bandwidth of 3.9 GHz, an analysis frequency up to 20 GHz, and a frequency resolution of 15 MHz are demonstrated, respectively.

## 1. Introduction

Microwave measurement plays an important role in the field of radiofrequency (RF) engineering. The quality of measurement often affects the performance of the RF systems [1], [2]. Spectrum detection of the signal under test (SUT) is an important part of microwave measurement, which has important applications in telecommunications, radar, navigation, biomedical instruments, and astronomical research [1]-[5]. Time-frequency analysis is an important tool for detecting the spectrum and monitoring the time-frequency distribution of microwave signals over a period of time [6]. Generally, Time-frequency analysis can obtain the following information: 1) The position of the SUT in the whole spectrum and the spectrum resources it occupies; 3) the discrimination of undesired frequency components other than the SUT; 3) real-time time-frequency distribution of the SUT [6], [7].

Conventional time-frequency analysis can be implemented by short-time Fourier transform (STFT), wavelet transform (WT), and Wigner-Ville distribution (WVD), among which STFT and WT are linear time-frequency analysis methods and WVD is a nonlinear method [6]-[9]. Although these methods have been widely used for time-frequency analysis, the implementation in the electrical domain always requires sampling the signal using an analog-to-digital converter (ADC). One fact is that the difficulty of manufacturing ADCs with high accuracy and high sampling speed often limits the bandwidth and speed of time-frequency analysis [10]. In addition, digital signal processing (DSP) is performed after sampling, in which the processing speed is thus inefficient in tracking spectral changes beyond a few microseconds, and faces challenges in measuring signals with bandwidths beyond the sub-GHz range [10], [11]. Compressive sensing enables signal reconstruction with undersampling, thus reducing the amount of data transmission, which can be used to increase the bandwidth of time-frequency analysis, but complex algorithms also limit the real-time performance of time-frequency analysis [11]-[13]. Therefore, new methods are urgently needed to be sought to conduct time-frequency analysis.

Microwave photonics (MWP) is an emerging interdisciplinary field that focuses on the generation, processing, control, and measurement of microwave signals, taking the advantages of large bandwidth, high frequency, good tunability, and immunity to electromagnetic interference, offered by modern optics [14]-[17]. It has been proved that photonics-assisted microwave measurement can break through the electrical bottleneck encountered in the electrical domain. To date, however, only a few studies using all-optical methods for time-frequency analysis of microwave signals have been proposed, among which the all-optical STFT is an effective method for time-frequency analysis. In [18], the SUT is modulated to a cluster of chirped optical pulses, a cascaded linear chirped fiber Bragg grating (LCFBG) array is used to extract specific frequencies and separate them from the time domain. In [19], a real-time time-frequency analysis scheme is proposed, where the SUT is modulated onto a cluster of optical pulses and then added to a section of dispersion-compensated fiber (DCF). After transmission in the DCF, the optical pulses of different frequencies are separated in the time domain, where the frequency resolution of the time-frequency analysis can reach 340 MHz. Bandwidth-magnified electrical-to-optical conversion is used in [20], which greatly reduces the dispersion requirement for high-frequency resolution all-optical STFT, as well as the transmission delay. The frequency resolution for time-frequency analysis is improved to 60 MHz. However, in all of the above work, the resolution of the time-frequency analysis is limited to tens of megahertz, which is difficult to meet the application scenarios that require higher frequency resolution.

Stimulated Brillouin scattering (SBS) is a typical nonlinear effect that is caused by the acousto-optic interaction. The Brillouin gain spectrum produced by the SBS effect can be used to amplify the optical signal in a narrow frequency band and has attracted great attentions in the past few years [21]-[23]. The frequency-to-time-mapping (FTTM) realized by the SBS effect is an effective solution to achieve a good frequency resolution in microwave frequency measurement [24]-[27]. We used the SBS-FTTM-based method to achieve microwave frequency measurements with a measurement error of no more than 1 MHz and a frequency resolution of about 20 MHz in [26]. The frequency resolution was further improved to less than 10 MHz based on the reduced SBS gain spectrum [27]. Based on our previous work on improving the frequency resolution of microwave frequency measurement using the SBS effect, we try to implement time-frequency analysis of SUTs using the SBS-FTTM-based method to effectively improve the analysis resolution. In addition, the bandwidth of the previously reported photonics-based methods is limited to 2.43 GHz, which is mainly limited by the dispersive medium. Thus, It is also desirable that new methods are deployed to increase the analysis bandwidth.

In this paper, two types of time-frequency analysis methods, based on a parallel SBS link or a time-division SBS link, are proposed for the time-frequency analysis of microwave signals. The parallel SBS link can be utilized to perform real-time and high-speed time-frequency analysis of microwave signal, which provides a promising solution for real-time time-frequency analysis, especially when it is combined with the photonic integration technique. A simulation is made to verify its feasibility. The time-division SBS link has a simpler and reconfigurable structure, which can realize an ultra-high-resolution time-frequency analysis for periodic signals using time segmentation and accumulation technique. An experiment is performed. The measurement bandwidth and frequency can be easily reconfigured by adjusting the system parameters, enabling time-frequency analysis of periodic microwave signals with an ultra-high frequency resolution of around 15 MHz. Furthermore, by using the time-frequency analysis based on the SBS effect, the bandwidth of the analysis can also be greatly increased compared with the previously reported methods.

## 2. Principle

FTTM based on the SBS effect is the basic principle of the time-frequency analysis method proposed in this paper. Based on the SBS-based FTTM structure, two types of time-frequency analysis methods are proposed for different application scenarios. In the following analysis, the FTTM based on the SBS effect is introduced first, followed by two proposed time-frequency analysis links based on it.

*2.1 Principle of the FTTM Based on the SBS effect*

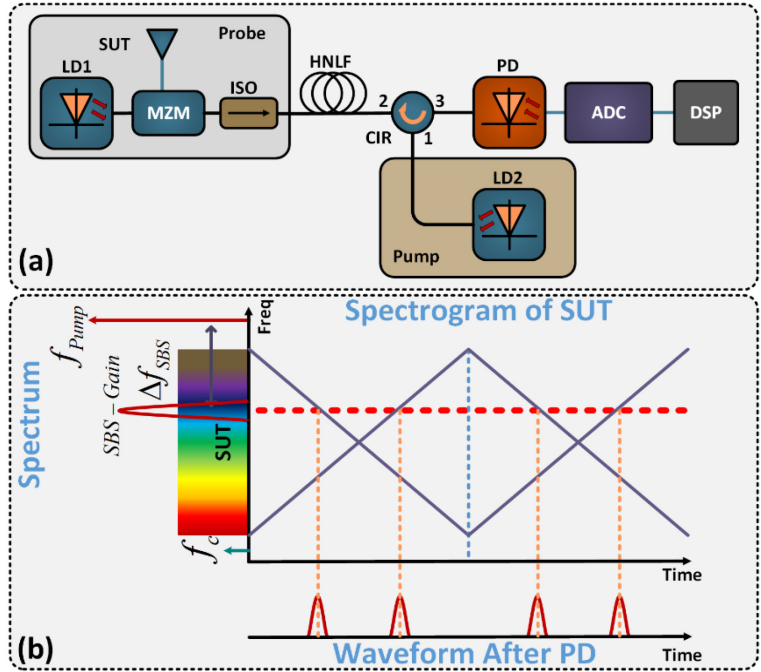

Fig. 1. (a) The basic structure for the FTTM based on the SBS effect, (b) the principle of mapping the SUT to electrical pulses. LD, laser diode; SUT, signal under test; MZM, Mach-Zehnder modulator; ISO, isolator; HNLF, highly nonlinear fiber; CIR, circulator; PD, photodetector.

The schematic diagram of the FTTM based on the SBS effect is shown in Fig. 1(a). An optical signal with a center frequency at $f_c$ from a laser diode (LD1) is sent to a Mach-Zehnder modulator (MZM), which is driven by the SUT. The optical signal from the MZM is sent to a section of highly nonlinear fiber (HNLF) via an isolator as the probe wave. The optical signal from LD2 is reversely injected into the HNLF as the pump wave. When the pump power reaches a certain value, the SBS effect is excited.

As shown in Fig. 1(b), when the pump wave from LD2 has a fixed frequency at $f_{pump}$, an SBS gain at $f_{SBS\text{-}Gain}=f_{pump}-f_{SBS}$ is generated, where $f_{SBS}$ is the Brillouin frequency shift. If the SUT is a dual-chirp wideband microwave signal as shown in Fig. 2(b), the optical sideband of the SUT from the MZM also scans over time, i.e., the position of the optical sideband varies with the RF signal at any instant. Therefore, when the optical sideband of SUT is scanned to the position of the SBS gain, it will be amplified. After the amplified optical is detected in a photodetector (PD), an electrical pulse is generated.

In the FTTM process, because the carrier frequencies of the two LDs remain unchanged, the frequency difference $f_1=f_{SBS\text{-}Gain}-f_c$ between the SBS gain and the optical carrier of LD1 remains unchanged. The generated pulses in Fig. 2(b) actually indicate that the SUT has a frequency component of $f_1$ at the time of pulse center, which means that one SBS-based FTTM structure can measure a specific frequency of the SUT. In addition, the intensity of the frequency component has a positive correlation with the amplitude of the generated electrical pulse.

*2.2 Time-frequency Analysis Using Parallel SBS Links*

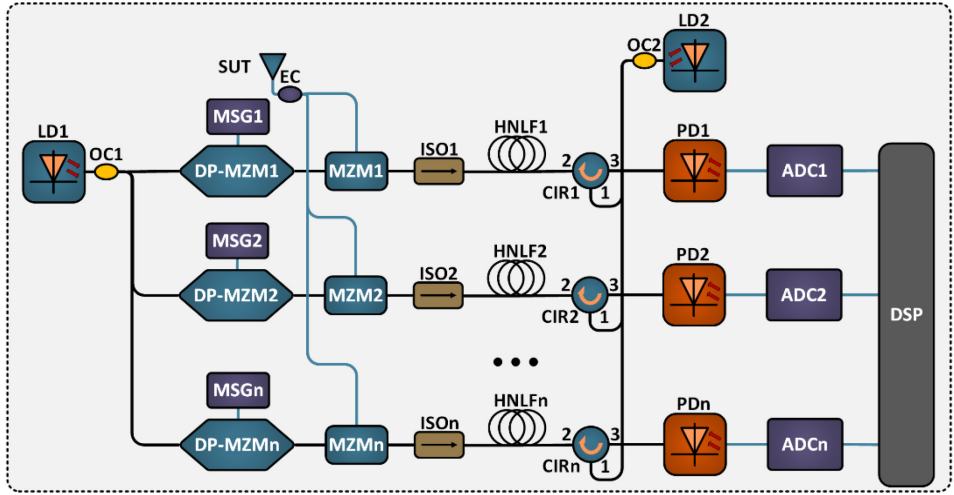

Fig. 2. Schematic diagram of the time-frequency analysis using multiple FTTM-based SBS links. LD, laser diode; OC, optical coupler; DP-MZM, dual-parallel Mach-Zehnder modulator; MZM, Mach-Zehnder modulator; SUT, signal under test; EC, electrical coupler; ISO, isolator; HNLF, highly nonlinear fiber; CIR, circulator; PD, photodetector; ADC, analog-to-digital convertor; DSP, digital signal processing.

By paralleling multiple FTTM-based SBS links shown in Fig. 1(a) and measuring a specific frequency with one parallel branch, real-time time-frequency analysis can be implemented. The schematic diagram of the time-frequency analysis using multiple FTTM-based SBS links is shown in Fig. 2. Compared with Fig. 1(a), a dual-parallel Mach-Zehnder modulator (DP-MZM) is inserted immediately after the LD in each branch, which is used to shift the frequency of the optical carrier sent to the followed MZM by carrier-suppressed single-sideband (CS-SSB) modulation using an electrical signal from a microwave signal generator (MSG). Assuming the frequencies applied on the MZMs are a series of discrete frequencies starting from 0 and spaced at $\Delta f$, the first channel can measure the signal frequency at $f_1$, and the frequencies that other channels can measure increase or decrease at intervals of $\Delta f$, depending on the direction of the CS-SSB modulation. In fact, because the SBS gain spectrum has a certain bandwidth, the frequency that each channel can measure is a frequency range around $f_1\pm n\Delta f$. Thus, the frequency resolution of the time-frequency analysis is determined by the frequency interval $\Delta f$. The smaller the frequency interval, the higher the frequency resolution of the time-frequency analysis. When $\Delta f$ is smaller than the SBS gain bandwidth, the frequency resolution will also be limited by the SBS bandwidth. In addition, the bandwidth of the SUT that can be analyzed can be extended by adding more branches in the parallel structure.

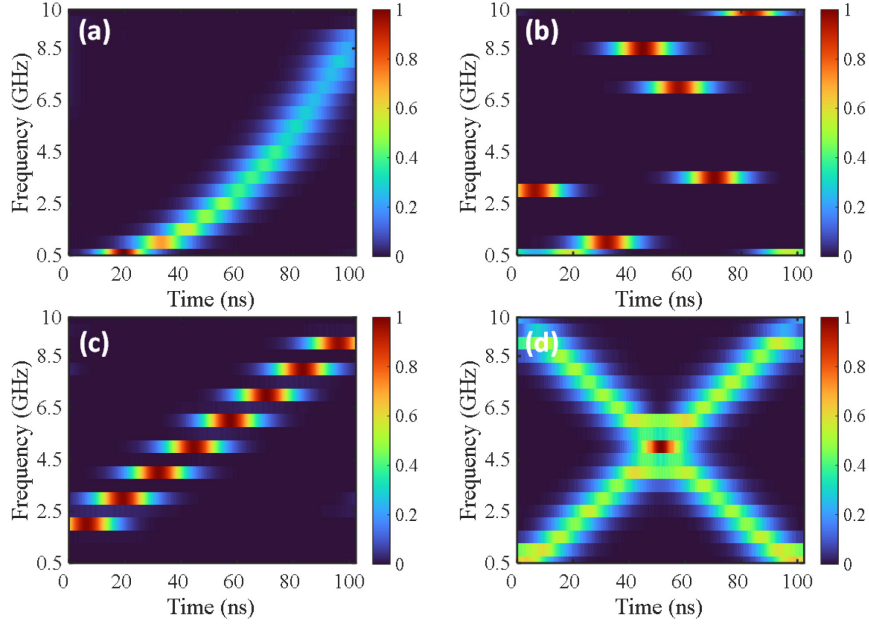

Fig. 3. Simulation results of the time-frequency diagrams of (a) non-linearly frequency-modulated signal, (b) frequency-hopping signal, (c) step frequency signal, and (d) dual-chirp linearly frequency-modulated microwave signal.

A simulation based on the setup shown in Fig. 2 is performed. Twenty parallel branches are employed to perform the time-frequency analysis. The frequency resolution is set to 500 MHz by setting the frequency step of the MSGs to 500 MHz. In this case, time-frequency analysis can be implemented from 0 to 10 GHz, while the time-frequency diagram of the SUT can be obtained by summarizing the electrical signals from each branch. Non-linearly frequency-modulated (NLFM) signal, frequency-hopping (FH) signal, step frequency (SF) signal, and dual-chirp linearly frequency-modulated (LFM) signal are chosen as the SUTs. The time-frequency diagram of an NLFM signal with a bandwidth of 9 GHz is shown in Fig. 3(a), which illustrates that the intensity is higher at lower frequencies, which is consistent with the result by using STFT. Fig. 3(b) and (c) show the time-frequency diagrams of the FH signal and the SF signal. For signals with discrete frequencies, when the frequencies of the signal are aligned with the SBS gain spectra, the time-frequency analysis will have a better result, which is actually due to the limitation of the resolution. From Fig. 3(d), the dual-chirp LFM signal is well time-frequency analyzed, which indicates that signals with multiple frequencies at a moment can be well detected by the parallel SBS link.

Simulation results indicate that the proposed parallel SBS link can perform time-frequency analysis of SUT in real time. Although this structure is complex, it will have a wide range of applications with its real-time analysis capability. In practice, parallel SBS structures with higher resolution and larger bandwidth can be designed according to application scenarios. The only problem to achieve a higher resolution and a larger bandwidth is the complexity of the system when more parallel links are incorporated. However, when such systems are integrated on chip based on integrated microwave photonics [31], [32], a better trade-off can be found between the system complexity and resolution and its practicability [5], [24], [28].

*2.3 Time-frequency Analysis Using Time-Division SBS Link*

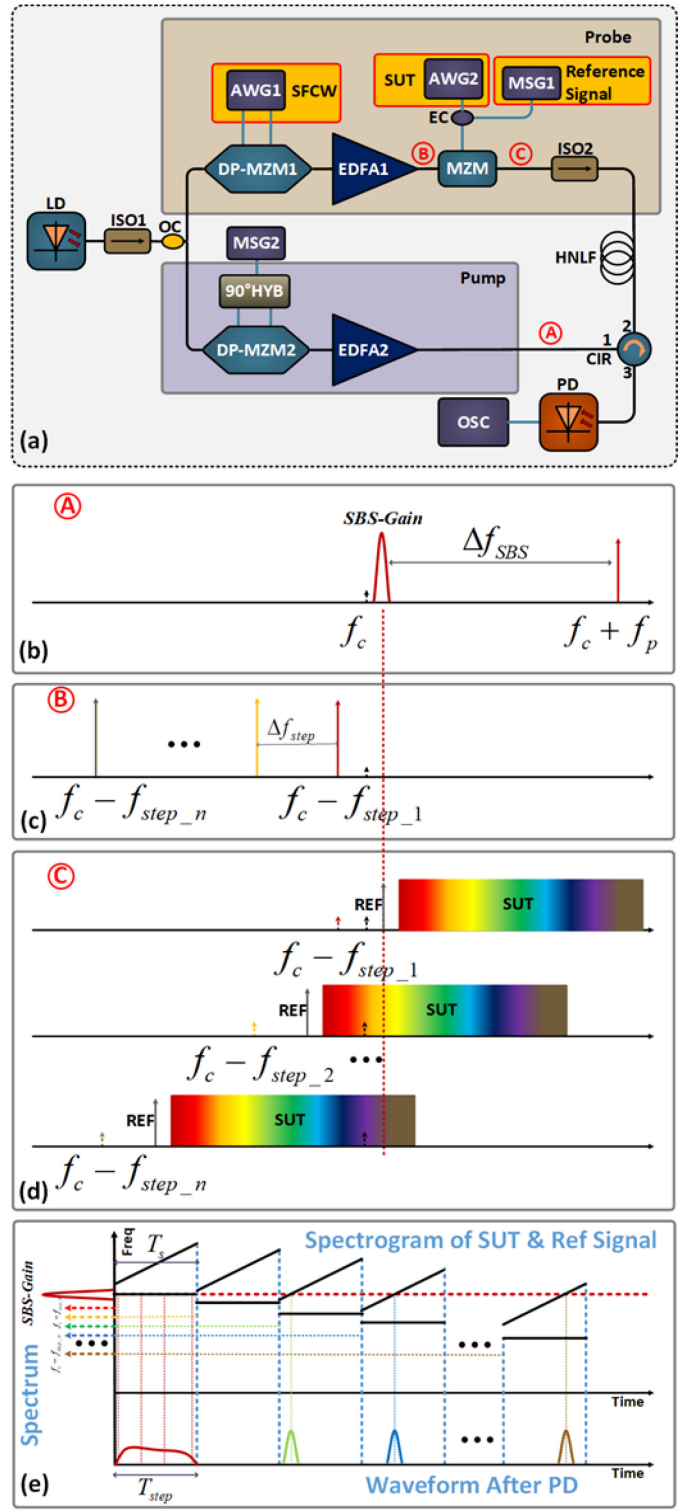

Fig. 4. Time-frequency analysis using the time-division SBS link. (a) Structure of the time-division SBS link, (b) spectrum at point A in the link, (c) spectrum at point B in the link, (d) spectrum at point C in the link, (e) the principle of mapping the SUT and the reference signal to electrical pulses. LD, laser diode; ISO, isolator; OC, optical coupler; DP-MZM, dual-parallel Mach-Zehnder modulator; EDFA, erbium-doped fiber amplifier; AWG, arbitrary waveform generator; MSG, microwave signal generator; 90°

HYB, 90° electrical hybrid coupler; MZM, Mach-Zehnder modulator; SFCW, step frequency continuous-wave; SUT, signal under test; EC, electrical coupler; HNLF, highly nonlinear fiber; CIR, circulator; PD, photodetector; OSC, oscilloscope.

The schematic diagram of time-division SBS links is shown in Fig. 4(a). An optical carrier with a frequency of $f_c$ generated from an LD is divided into two parts via an optical coupler (OC). In the lower path, the optical carrier is carrier-suppressed upper single-sideband (CS-USSB) modulated by a single-tone RF signal with a frequency of $f_p$ at DP-MZM2. After being amplified in an erbium-doped fiber amplifier (EDFA2), the output of DP-MZM2 is launched into a spool of HNLF via an optical circulator as the pump wave, where it interacts with the counter-propagating wave from the upper path and will generate an SBS gain with its frequency centered at $f_{SBS\text{-}Gain} = f_c + f_p - f_{SBS}$, as shown in Fig. 4(b). In the upper path, the optical carrier is carrier-suppressed lower single-sideband (CS-LSSB) modulated by a step frequency continuous-wave (SFCW) signal at DPMZM1, with its spectrum shown in Fig. 4(c), where $f_{step\_1}$ is the initial frequency of the SFCW signal, $\Delta f_{step}$ is the step frequency interval, $T_{step}$ is the step period, and $n$ is the total number of steps in a period. Then, the output of DP-MZM1 serves as a new optical carrier after being amplified by EDFA1 and is injected into an MZM. The SFCW optical carrier is carrier-suppressed double-sideband (CS-DSB) modulated at the null-biased MZM by the single-tone reference signal and the SUT with a period of $T_s$. The first-order optical sidebands of the modulated broadband periodic signal jump with the carrier and the signal spectra in different step periods are shown in Fig. 4(d). The negative sidebands that do not interact with the SBS gain during the scanning process are not given in Fig. 4(d) and (e). The output of the MZM that functions as the probe wave is injected into the HNLF through an optical isolator. In the HNLF, the probe wave is amplified by the fixed SBS gain. Then, the optical signal from the HNLF is detected in a PD and monitored by an oscilloscope (OSC).

To implement the time-frequency analysis of the SUT, it must be ensured that the fixed SBS gain can conduct a complete scan of the SUT in each step period to determine whether and when the SUT has a frequency at the frequency point corresponding to the SBS gain. Therefore, the step period must be an integer multiple of the SUT period,

$$T_{step} = m \cdot T_s, m \in N^+ . \tag{1}$$

Fig. 4(e) is the principle of mapping the SUT and the reference to electrical pulses in the specific case where $m=1$. As can be seen, electrical pulses at different time will be generated in different periods during the scanning process when the positive sidebands are detected by the fixed SBS gain spectrum. The time-frequency diagram of the SUT with a period of $T_s$ is obtained by detecting different frequency components at a different time in different step periods. The pulses generated in the time domain through the FTTM in different step periods are segmented and recombined to obtain the time-frequency diagram of the SUT. As shown in Fig. 4(a), a known reference signal is applied to the system along with the SUT, which is mainly used to provide a label of time and frequency during the recombination.

If an electrical pulse is generated from the PD at a given point in a step period when the step period is set to the same as the SUT period, it signifies that the SUT has the frequency component that the link can measure in this step period. The first step corresponding to $f_{step\_1}$ can measure the following frequency component in the SUT,

$$f_1 = f_{step\_1} - f_c + f_{SBS-Gain} . \tag{2}$$

The $n$-th step corresponding to $f_{step\_n}$ can measure the following frequency component,

$$f_n = f_{step\_1} - f_c + (n-1) \cdot \Delta f_{step} + f_{SBS-Gain}. \tag{3}$$

In fact, what the system can measure in a single period is not a single frequency, but a frequency range highly related to the SBS bandwidth. Thus, the time resolution of the time-frequency analysis is determined by SBS gain bandwidth. When all pulses of different periods are analyzed together, the frequency resolution of the time-frequency analysis is mainly determined by $\Delta f_{step}$. However, when $\Delta f_{step}$ is smaller than the SBS gain bandwidth, the frequency resolution will also be limited by the SBS bandwidth. Therefore, the resolution of the time-frequency analysis is jointly determined by the SBS gain bandwidth and $\Delta f_{step}$. Commonly, the SBS gain bandwidth is roughly a fixed value, unless some bandwidth reduction methods are employed [27]. Therefore, when the SBS bandwidth is fixed, i.e., the time resolution is fixed, $\Delta f_{step}$ is the key factor to determine the resolution of the time-frequency analysis. The smaller the frequency step $\Delta f_{step}$, the higher the frequency resolution of the system. When the analysis bandwidth of the system is fixed, more periods are required to achieve a better frequency resolution by employing a smaller $\Delta f_{step}$, which will increase the analysis time of the system. When the frequency step $\Delta f_{step}$ is fixed, the analysis bandwidth of the system can be reconstructed by changing the number of steps of the SFCW signal. The more SFCW steps, the greater the bandwidth of the time-frequency analysis. In addition, the frequency band that can be time-frequency analyzed can be changed by moving the SBS gain via changing $f_p$.

In a word, the parameters of the system are controlled according to the application scenarios to achieve the purpose of optimizing the analysis time and the analysis resolution. The analysis resolution and bandwidth of the proposed method can be much better than the previously reported methods at a cost of a relatively longer analysis time. The only limitation of the system is that only a periodic signal with a known period can be measured and it is necessary to accumulate for many periods to achieve a good resolution. However, the method of signal interception and pulse replication [29], [30] can be used to extend this method to a more generalized method to achieve the measurement of other kinds of signal, such as pulse signals.

## 3. Experimental results and discussion

### 3.1 Experimental Setup

An experiment based on the setup shown in Fig. 3(a) is performed to verify the time-division SBS link for the time-frequency analysis of periodic SUTs. A 15-dBm optical carrier with a frequency of 193.215 THz from the LD (ID Photonics CobriteDX1-1-C-H01-FA) is divided into two paths via an OC. To generate an SFCW optical carrier, one output of the OC is CS-LSSB modulated at DP-MZM1 (Fujitsu FTM 7961EX) by a designed SFCW electrical signal from AWG1 (Keysight M8190A) in the upper path. Here, to achieve the CS-LSSB modulation, two SFCW electrical signals with a 90° phase difference generated from AWG1 are inject into the two RF ports of DPMZM1, and the two sub-MZMs of DP-MZM1 are both null-biased and the main-MZM is biased to introduce a 90° phase shift. Subsequently, the output from DPMZM1, serving as a new SFCW optical carrier, is injected into the MZM (Fujitsu, FTM 7938 EZ) after being amplified by EDFA1 (Max-ray EDFA-PA-35-B). The SFCW optical carrier is CS-DSB modulated at the null-biased MZM by a combined signal from the electrical coupler (EC, Narda 4456-2), which includes the SUT generated from AWG2 (Keysight M8195A) and the single-tone reference signal generated from MSG1 (HP 83752B). The optical signal from the MZM is injected into a section of 25.2-km single-mode fiber (SMF) through an optical isolator. In the experiment, the HNLF is replaced by the SMF due to the lack of the HNLF in our laboratory. In the lower path, the optical carrier from the OC is CS-USSB modulated at DP-MZM2 (Fujitsu FTM 7961EX) by an electrical signal from MSG2 (Agilent 83630B). Here, to achieve the CS-USSB modulation, a 90° electrical hybrid coupler (90° HYB, Narda 4065) is used between the MSG and DP-MZM2, and the two sub-MZMs of the DP-MZM2 are both null-biased and the main-MZM is biased to introduce a 90° phase shift. The output of DP-MZM2 is amplified by EDFA2 (Amonics AEDFA-PA-35-B-FA) and then launched into the SMF via an optical circulator as the pump wave, where it interacts with the counter-propagating wave from the upper branch. The optical signal from the SMF is detected in a PD (Nortel PP-10G) and monitored by the OSC(R&S RTO2032). The data collected by the OSC is extracted and processed to obtain the time-frequency analysis results of the SUT. Commonly, the reference can be set at the edge of the measurement bandwidth, and the SUT is preferred to have no frequency coincidence with the reference.

The data processing consists of the following steps: 1) finding the wide and high pulse generated by the reference signal; 2) using the wide pulse found in the first step as the starting point to extract the signal with the same length as the period of SFCW signal; 3) dividing the signal into multiple segments in time, and the number of segments is the same as the number of steps of the SFCW signal; 4) the multi-segment signals are formed into a matrix, and the time-frequency diagram of the system is obtained through the matrix.

*3.2 Frequency and Bandwidth Tunability*

The initial frequency of the SFCW signal generated from AWG1 is set to 50 MHz, the step interval is set to 5 MHz, and the step period is set to 2 μs with 781 steps in total. The SBS frequency shift of the system is measured to be around 10.8 GHz. The frequency of single-tone RF signal from MSG2 is set to 10.85 GHz. Under these conditions, it is expected that the system is capable of performing time-frequency analysis of the SUT with a bandwidth ranging from 100 MHz to 4 GHz. So, an LFM signal with a bandwidth ranging from 100 MHz to 4 GHz and a period of 2 μs generated from AWG2 is chosen as the SUT. The reference frequency is set to 100 MHz. Fig. 5(a) shows the time-frequency diagram of the LFM signal, which shows good consistency with the parameters we set. In processing the data, the first period corresponding to the reference signal is removed in generating the time-frequency diagram.

In the experiment, limited by the bandwidth of AWG1, the time-frequency analysis frequency range is demonstrated from 100 MHz to 4 GHz. The bandwidth is difficult to be further increased due to the limitation of the instrument. However, the operating frequency can be easily adjusted by simply tuning the wavelength from the pump branch, which can be implemented by changing the RF frequency from MSG2. Due to the lack of a 90° HYB that can work above 20 GHz, in the experiment, the frequency tunability is verified by replacing the frequency-shift module in the lower path with another independent LD (ID Photonics CobriteDX1-1-C-H01-FA), which is used as a pump source with flexible and adjustable wavelength. The two LDs work independently and their wavelengths are calibrated using the reference signal.

The step period, step interval, and step number of the SFCW signal are set to 2 μs, 25 MHz, and 157, respectively. The frequency differences between the two LDs are set to 14.85, 18.85, 24.85, and 28.85 GHz, respectively. Under these circumstances, it is expected that the frequency bands that can be analyzed are 4.1 to 8 GHz, 8.1 to 12 GHz, 12.1 to 16 GHz, and 16.1 to 20 GHz, accordingly. LFM signals set in the above frequency bands are time-frequency analyzed by the system, with the results shown in Fig. 5(b) to (e). As can be seen, the time-division SBS link can perform the time-frequency analysis of the SUT in different frequency bands simply by altering the frequency of the pump wave. It is also observed that the intensity of the result shown in Fig.5(e) is slightly lower than the other curves, which is mainly caused by the relatively larger insertion loss of the cables and electrical devices used to connect AWG2 and the MZM in the higher frequency band.

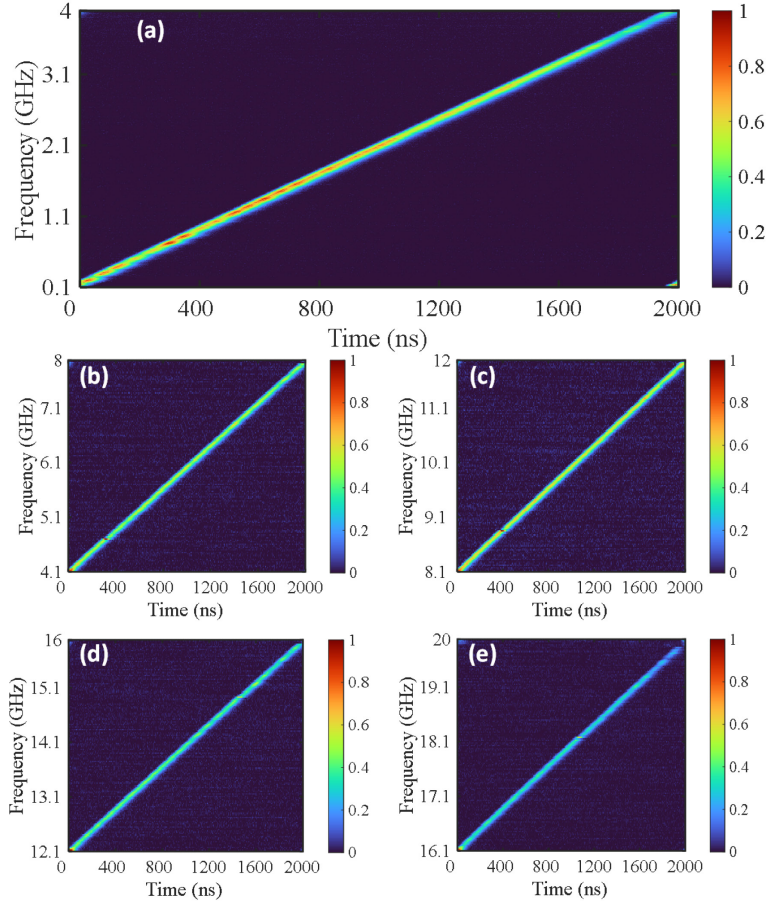

Fig. 5. Time-frequency diagrams of the LFM signals from (a) 0.1 to 4 GHz, (b) 4.1 to 8 GHz, (c) 8.1 to 12 GHz, (d) 12.1 to 16 GHz, and (e) 16.1 to 20 GHz.

*3.3 Analysis Resolution*

To demonstrate the analysis resolution of the system, SFCW signals with different step intervals of 100, 50, 25, 10, and 5 MHz are used to analyze the same SUT. Because the tunability is not the focus of the following parts, in all the following experiments, the system is set up as shown in Fig. 4(a) to have the time-frequency analysis capability from 0.1 to 4 GHz. In the experiment, no bandwidth reduction methods are employed, so we mainly focus on the frequency resolution of the time-frequency analysis.

The start and end frequencies of the SFCW signal are always set to 50 MHz and 3.95 GHz, respectively, regardless of the step interval of the SFCW signal. An LFM with a bandwidth ranging from 0.19 to 4 GHz is used as the SUT. Fig. 6 shows the time-frequency diagrams of the SUT under different step intervals. As indicated from Fig. 6, with the decrease of the step interval, the frequency resolution of the time-frequency diagram gets better and better. Furthermore, when the step interval of the SFCW signal is around or less than the SBS gain bandwidth (about 20 MHz in the experiment), the resolution of the diagram is mainly limited by the SBS gain bandwidth. That is why Fig. (c) to (e) are very similar under the coordinates and scales in the figure.

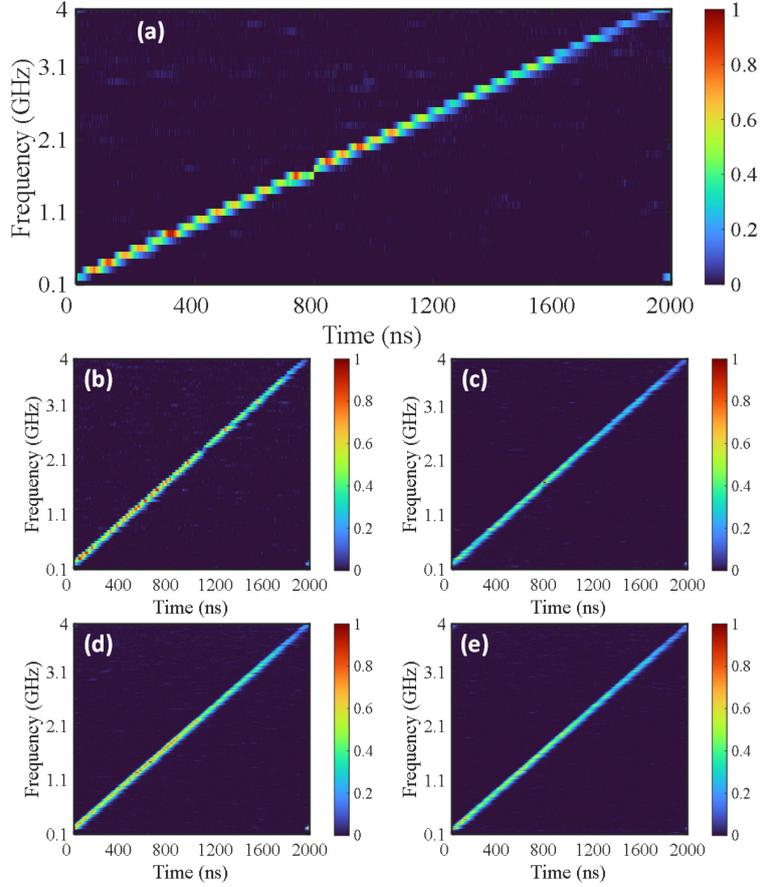

Fig. 6. Time-frequency diagrams of the LFM signals from 0.19 to 4 GHz when the step interval of the SFCW signal is (a) 100 MHz, (b) 50 MHz, (c) 25 MHz, (d) 10 MHz, and (e) 5 MHz.

Then, a two-tone test is further carried out to show the resolution more intuitively. In this study, a two-tone signal at 3 and 3.025 GHz are chosen as the SUTs. Time-frequency analysis of the two-tone signals is performed using SFCW signals with step intervals of 25 and 5 MHz, respectively. As shown in Fig. 7(a) and (b), when the step interval of the SFCW signal is 25 MHz, the single-tone signals at 3 and 3.025 GHz are mapped to a straight line in the time-frequency diagram. However, when the two single-tone signals are combined as a two-tone signal and then analyzed using the proposed system under the same step interval of 25 MHz, the two lines corresponding to the two frequencies in the time-frequency diagram fit together, as shown in Fig. 7(c). To clearly distinguish the curves representing the 3- and 3.025-GHz signals in the time-frequency diagram, the step interval of the SFCW signal is decreased to 5 MHz and the corresponding time-frequency diagram is shown in Fig. 7(d). In this case, as can be seen from the inset, the two signals with a frequency difference of 25 MHz can be clearly distinguished.

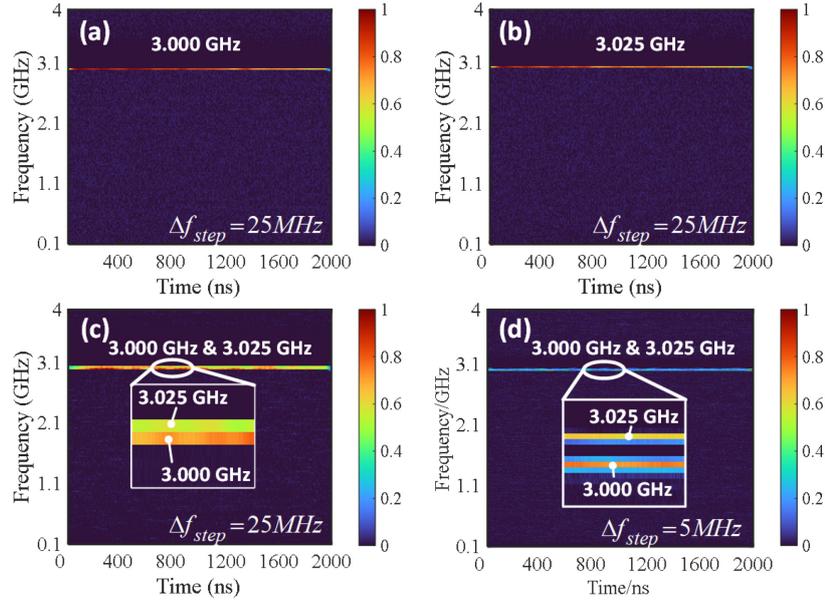

Fig. 7. Time-frequency diagrams of a single-tone signal at (a) 3 GHz and (b) 3.025 GHz when the step interval is 25 MHz. Time-frequency diagrams of a two-tone signal at 3 and 3.025 GHz when the step interval is (c) 25 MHz and (d) 5 MHz. The insets in (c) and (d) are the zoom-in views of the curves.

It is also observed that when the step interval of the SFCW signal is 5 MHz, there are relatively weak power horizontal lines around the lines representing the SUT. The unwanted horizontal lines appear because the SBS gain spectrum has a certain bandwidth. Therefore, not only the period of the SFCW signal that makes SUT closest to the center of the SBS gain spectrum will perform the FTTM, but also its adjacent periods will perform the FTTM when the step interval of the SFCW signal is very small. That is because it also makes the SUT within the range of the SBS gain spectrum with a relatively lower gain. Thus the power of the unwanted horizontal lines in the time-frequency curve is lower. Therefore, when the step interval of the SFCW signal is smaller than the SBS gain bandwidth, the frequency resolution will also be limited by the SBS bandwidth. When the frequency difference of the two-tone signal is further reduced, they can be still distinguished when the frequency difference is 15 MHz.

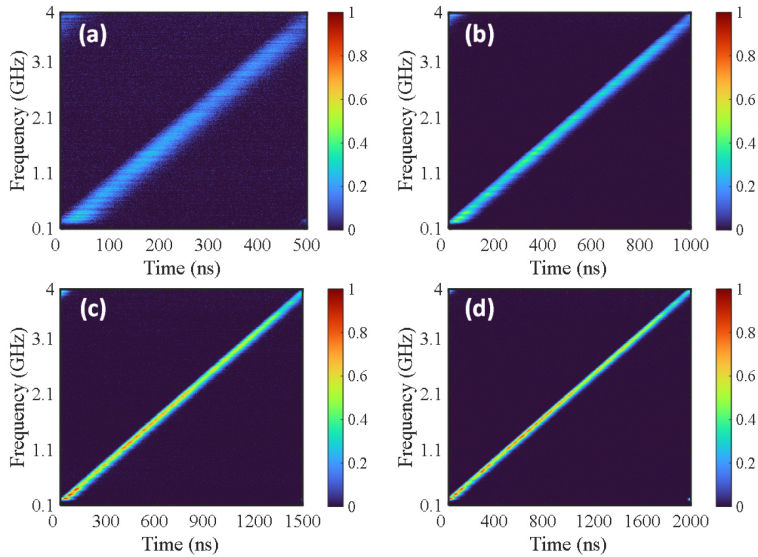

Fig. 8. Time-frequency diagrams of SUTs with different periods of (a) 500 ns, (b) 1000 ns, (c) 1500 ns, and (d) 2000 ns.

The periods of the SUT in the above experiments are all 2 μs. When the signal period changes, the period of the SFCW signal should be adjusted accordingly. The time-frequency diagrams of the LFM signals with different periods of 0.5, 1, 1.5, and 2 μs are shown in Fig. 8. As can be seen, the pulses generated by the FTTM have a certain time width around 90 ns, which is directly related to the time resolution of the time-frequency analysis. As discussed in Section II C, the time resolution is determined by the SBS gain bandwidth. Therefore, when the SUT has a short time length, the measurement time resolution must be improved by reducing the SBS gain spectrum width in order to realize the accurate analysis of short-time signals [27].

*3.4 Time-Frequency Analysis of Multi-Format Signals*

In order to verify the capability of the proposed time-frequency analysis system to any kind of signal, time-frequency analysis of multi-format signals is further studied. In the experiment, different kinds of wideband periodic signals, including NLFM signal, FH signal, SF signal, and dual-chirp LFM signal, are chosen as the SUTs. The step interval of the SFCW signal is set to 5 MHz. The period of each type of signal is 2 μs and the signal bandwidth covers the frequency range from 0.1 to 4 GHz. The results of the time-frequency analysis are shown in Fig. 9. As can be seen, the time-frequency relationships of the four kinds of signals are well constructed with good resolution, which confirms the capability of the system for the time-frequency analysis of multi-format signals.

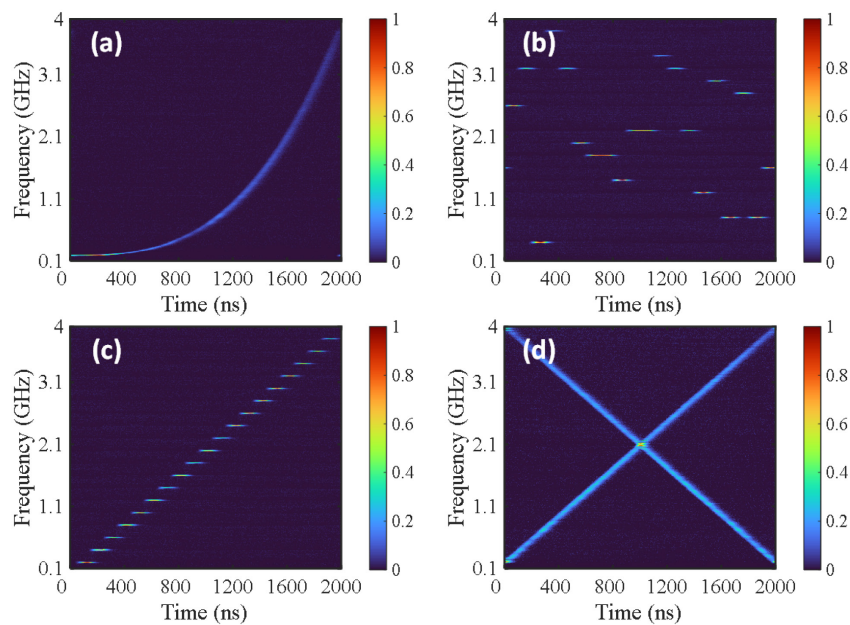

Fig. 9.  Time-frequency diagrams of (a) non-linearly frequency-modulated signal, (b) frequency-hopping signal, (c) step frequency signal, and (d) dual-chirp linearly frequency-modulated

*3.5 Periodic Relationships*

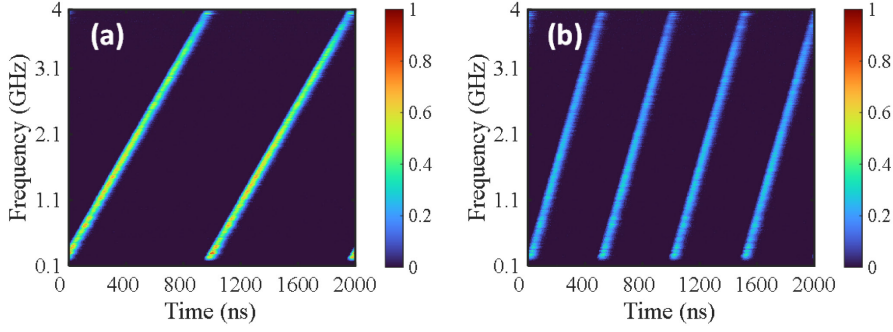

Fig. 10. Time-frequency diagram of the SUT when the step period of the SFCW signal is (a) twice and (b) four times that of the SUT period.

As mentioned in Section II, the step period of the SFCW signal should be an integer multiple of the SUT period. In the above experiments, we set the two periods to be equal for convenience. When the SUT period is short, a long-period SFCW signal with an integer multiple of the SUT period can be used to reduce the difficulty of generating the SFCW signal. Two experiments are carried out by setting the step period of the SFCW signal (2 μs) to be twice (1 μs) or four times (0.5 μs) that of the SUT period. Under these conditions, LFM signals with a different period are time-frequency analyzed using the same SFCW signal, with the results shown in Fig. 10. It can be seen that the proposed system can also perform time-frequency analysis of signals with short periods without changing the parameters of the SFCW signal. Fig. 10 also illustrates that in such a case, signals of multiple signal periods are time-frequency analyzed in one SFCW signal period. However, the time resolution of the time-frequency analysis using this method is still bounded by the pulse width of the FTTM, i.e., the SBS gain bandwidth.

*3.6 Comparison Between Different Methods*

TABLE I

COMPARISON OF DIFFERENT PHOTONICS-BASED TIME-FREQUENCY ANALYSIS METHODS

|  | Specific Techniques/Devices Used | Structure Complexity | Demonstrated Bandwidth | Frequency Resolution | Demonstrated Frequency | Real-time Availability | Reconfigurability |
|---|---|---|---|---|---|---|---|
| Ref.[18] | LCFBG Array | Middle | Unspecified [1] | | 3 GHz | Yes | Low |
| Ref.[19] | DCF | Low | 2.43 GHz | 340 MHz | 2.43 GHz | Yes | Low |
| Ref.[20] | Time lens and DCF | High | 1.98 GHz | 60 MHz | 2.58 GHz | Yes | Low |
| Parallel SBS link | SBS and FTTM | High | / [2] | | | Yes | Middle |
| Time-division SBS link | SBS and FTTM | Middle | 3.9 GHz [3] | 15 MHz [4] | 20 GHz [3] | No | High |

[1] Only the analysis of the two electrical signals with a frequency of 1.5 GHz and 3GHz was performed;

[2] Not experimentally demonstrated, and determined by the number of parallel structures and the frequency shift of each branch;

[3] These parameters are limited by the bandwidth of the SFCW signal and the pump frequency, which can be further increased;

[4] Limited by the SBS gain bandwidth and can be further improved by SBS gain bandwidth reduction [27].

A comparison between the proposed time-frequency analysis method and the previously reported ones is made and given in Table I. As can be seen, the reported methods are all based on dispersion devices, such as the LCFBG and the DCF. However, the dispersion devices are hard to be reconstructed in real-time in multiple dimensions, which severely constrains the reconfigurability of these methods [18]-[20]. In comparison, the proposed time-division SBS link for time-frequency analysis can be reconstructed in multiple dimensions (bandwidth, frequency, and frequency resolution) by simply changing the input SFCW signal and the wavelength of the pump wave. As shown in Table I, thanks to the narrow and flexible gain spectrum enabled by the SBS effect, the proposed time-division SBS link has the best operating frequency, bandwidth, and frequency resolution over the reported results [18]-[20]. The frequency resolution is greatly improved to 15 MHz, which can be further improved by SBS gain bandwidth reduction [27]. The bandwidth and frequency of the time-frequency analysis in the experiment are limited mainly by working bandwidths of the two AWGs, which can also be further increased using better instruments. Compared with the time-frequency analysis methods in [18]-[20], the major limitation of the demonstrated time-division SBS link is that it cannot perform real-time time-frequency analysis and can only be used to analyze periodic signals. However, the method of signal interception and pulse replication [29], [30] can be used to extend this method to a more generalized method to achieve the measurement of other kinds of signal, such as pulse signals.

## 4. Conclusions

In this paper, two types of time-frequency analysis methods, based on a parallel SBS link or a time-division SBS link, are proposed for the time-frequency analysis of microwave signals. The parallel SBS link can be utilized to perform real-time and high-speed time-frequency analysis of microwave signal, which is analyzed and verified via a numerical simulation. The time-division SBS link can realize an ultra-high-resolution time-frequency analysis for periodic signals using the time segmentation and accumulation technique, which is analyzed and experimentally verified. Different kinds of the periodic SUTs are analyzed using the system in the frequency range from 0.1 to 20 GHz with the best resolution of around 15 MHz. The key significance of the work is that, for the first time, the SBS-based FTTM structure is applied to the time-frequency analysis system, which achieves the best operating frequency, bandwidth, and frequency resolution over the reported results. The proposed method provides a promising new solution for time-frequency analysis, which may find applications in advanced microwave measurement systems.

**Funding**

This work was supported in part by the National Natural Science Foundation of China under Grant 61971193, in part by the Natural Science Foundation of Shanghai under Grant 20ZR1416100, in part by the Open Fund of State Key Laboratory of Advanced Optical Communication Systems and Networks, Peking University, China, under Grant 2020GZKF005, and in part by the Science and Technology Commission of Shanghai Municipality under Grant 18DZ2270800.

**Conflicts of interest**

The authors declare no conflicts of interest.

## References

[1] X. Zou, B. Lu, W. Pan, L. Yan, A. Stöhr, and J. Yao, "Photonics for microwave measurements," *Laser Photon. Rev.*, vol. 10, no. 5, pp. 711–734, Sept. 2016.

[2] S. Pan and J. Yao, "Photonics-Based Broadband Microwave Measurement," *J. Lightw. Technol.*, vol. 35, no. 16, pp. 3498–3513, Aug. 2017.

[3] M. A. Richards, *Fundamentals of Radar Signal Processing*. New York, NY, USA: McGraw-Hill Education, 2014.

[4] A. DeMartino, *Introduction to Modern EW Systems*. Norwood, MA, USA: Artech House, 2012.

[5] L. Romero Cortés, D. Onori, H. Guillet de Chatellus, M. Burla, and J. Azaña, "Towards on-chip photonic-assisted radio-frequency spectral measurement and monitoring," *Optica*, vol. 7, no. 5, pp. 434–447, May. 2020.

[6] L. Cohen, "Time-frequency distributions-a review," *Proc. IEEE*, vol. 77, no. 7, pp. 941–981, Jul. 1989.

[7] L. Cohen, "The scale representation," *IEEE Trans. Signal Process.*, vol. 41, no. 12, pp. 3275–3292, Dec. 1993.

[8] B. Boashash and P. Black, "An efficient real-time implementation of the Wigner-Ville distribution," *IEEE Trans. Acoust., Speech, Signal Process.*, vol. 35, no. 11, pp. 1611–1618, Nov. 1987.

[9] B. Boashash, "Estimating and interpreting the instantaneous frequency of a signal. I. Fundamentals," *Proc. IEEE*, vol. 80, no. 4, pp. 520–538, Apr. 1992.

[10] R. H. Walden, "Analog-to-digital converter survey and analysis," *IEEE J. Sel. Areas Commun.*, vol. 17, no. 4, pp. 539–550, Apr. 1999.

[11] E. J. Candes and M. B. Wakin, "An Introduction to Compressive Sampling," *IEEE Signal Process. Mag.*, vol. 25, no. 2, pp. 21–30, Mar. 2008.

[12] M. F. Duarte and Y. C. Eldar, "Structured Compressed Sensing: From Theory to Applications," *IEEE Trans. Signal Process*, vol. 59, no. 9, pp. 4053–4085, Sept. 2011.

[13] E. Sejdic, I. Orovic, and S. Stankovic, "Compressive sensing meets time-frequency: An overview of recent advances in time-frequency processing of sparse signals," *Digit Signal Process.*, vol. 77, pp. 22–35, Jun. 2018.

[14] A. J. Seeds and K. J. Williams, "Microwave photonics," *J. Lightw. Technol.*, vol. 24, no. 12, pp. 4628–4641, Dec. 2006.

[15] J. Capmany and D. Novak, "Microwave photonics combines two worlds," *Nat. Photon.*, vol. 1, no. 6, pp. 319–330, Jun. 2007.

[16] J. Yao, "Microwave photonics," *J. Lightw. Technol.*, vol. 27, no. 3, pp. 314–335, Feb. 2009.

[17] T. Berceli, and P. R. Herczfeld, "Microwave photonics-a historical perspective," *IEEE Trans. Microw. Theory Tech.*, vol. 58, no. 11, pp. 2992–3000, Nov. 2010.

[18] M. Li and J. Yao, "All-optical short-time fourier transform based on a temporal pulse-shaping system incorporating an array of cascaded linearly chirped fiber bragg gratings," *IEEE Photon. Technol. Lett.*, vol. 23, no. 20, pp. 1439–1441, Oct. 2011.

[19] S. R. Konatham, R. Maram, L. R. Cortés, J. H. Chang, L. Rusch, S. LaRochelle, H. G. de Chatellus and J. Azaña, "Real-time gap-free dynamic waveform spectral analysis with nanosecond resolutions through analog signal processing," *Nat. Commun.*, vol. 11, no. 1, Dec. 2020, Art. no. 3309.

[20] X. Xie, J. Li, F. Yin, K. Xu, and Y. Dai, "STFT Based on Bandwidth-Scaled Microwave Photonics," *J. Lightw. Technol.*, vol. 39, no. 6, pp. 1680–1687, Mar. 2021.

[21] B. Yang, X. Jin, H. Chi, X. Zhang, S. Zheng, S. Zou, H. Chen, E. Tangdiongga, and T. Koonen, "Optically tunable frequency-doubling Brillouin optoelectronic oscillator with carrier phase-shifted double sideband modulation," *IEEE Photon. Technol. Lett.*, vol. 24, no. 12, pp. 1051–1053, Jun. 2012.

[22] H. Peng, C. Zhang, X. Xie, T. Sun, P. Guo, X. Zhu, L. Zhu, W. Hu, and Z. Chen, "Tunable DC-60 GHz RF generation utilizing a dual-loop optoelectronic oscillator based on stimulated Brillouin scattering," *J. Lightw. Technol.*, vol. 33, no. 13, pp. 2707–2715, Jul. 2015.

[23] X. Yao, "Brillouin selective sideband amplification of microwave photonic signals," *IEEE Photon. Technol. Lett.*, vol. 10, no. 1, pp. 138–140, Jan. 1998.

[24] H. Jiang, D. Marpaung, M. Pagani, K. Vu, D. Choi, S. J. Madden, L. Yan, B. J. Eggleton, "Wide-range, high-precision multiple microwave frequency measurement using a chip-based photonic Brillouin filter," *Optica*, vol. 3, no. 1, pp. 30–34, Jan. 2016.

[25] S. Zheng, S. Ge, X. Zhang, H. Chi and X. Jin, "High-Resolution Multiple Microwave Frequency Measurement Based on Stimulated Brillouin Scattering," *IEEE Photon. Technol. Lett.*, vol. 24, no. 13, pp. 1115–1117, Jul. 2012.

[26] J. Liu, T. Shi, and Y. Chen, "High-Accuracy Multiple Microwave Frequency Measurement With Two-Step Accuracy Improvement Based on Stimulated Brillouin Scattering and Frequency-to-Time Mapping," *J. Lightw. Technol.*, vol. 39, no. 7, pp. 2023–2032, Apr. 2021.

[27] T. Shi and Y. Chen, "Multiple radio frequency measurements with an improved frequency resolution based on stimulated Brillouin scattering with a reduced gain bandwidth," *Opt. Lett.*, vol. 46, no. 14, pp. 3460–3463, Jul. 2021.

[28] D. Marpaung, B. Morrison, M. Pagani, R. Pant, D. Choi, B. Luther-Davies, S. J. Madden, B. J. Eggleton, "Low-power, chip-based stimulated Brillouin scattering microwave photonic filter with ultrahigh selectivity," *Optica*, vol. 2, no. 2, pp. 76–83, Feb. 2015.

[29] Z. Ding, F. Yang, J. Zhao, R. Wu, H. Cai, M. Wang, Y. Weng, Z. Zhao, "Photonic high-fidelity storage and Doppler frequency shift of broadband RF pulse signals," *Opt. Express*, vol. 27, no. 23, pp. 34359–34369, Nov. 2019.

[30] Z. Ding, F. Yang, J. Zhao, R. Wu, H. Cai, M. Wang, Y. Weng, Z. Zhao, "Multifunctional photonic broadband RF memory for complex electronic jamming," *Laser Phys. Lett.,* vol. 17, no. 11, Oct. 2020, Art. no. 116201.

[31] D. Marpaung, J. Yao, and J. Capmany, "Integrated microwave photonics," *Nat. Photon.*, vol. 13, no. 2, pp. 80–90, Feb. 2019.

[32] D. Marpaung, C. Roeloffzen, R. Heideman, A. Leinse, S. Sales, and J. Capmany, "Integrated microwave photonics," *Laser Photon. Rev.*, vol. 7, no. 4, pp. 506–538, Jul. 2013.